\documentclass[prb,aps,floats,floatsfix,preprint]{revtex4}
\usepackage{amsfonts}
\usepackage{amsmath}
\usepackage{amssymb}
\usepackage{color,graphicx}
\usepackage{float}

\begin{document}

\title{A Versatile Ultra-Stable Platform for Optical Multidimensional Fourier-Transform Spectroscopy}
\author{A.~D.~Bristow, D.~Karaiskaj, X.~Dai, T.~Zhang, C.~Carlsson, K.~R.~Hagen, R.~Jimenez and S.~T.~Cundiff}
\email{cundiffs@jila.colorado.edu}
\address{JILA, University of Colorado \& National Institute of Standards and
Technology, Boulder, Colorado 80309-0440, USA}

\begin{abstract}
The JILA Multidimensional Optical Nonlinear SpecTRometer
(JILA-MONSTR) is a robust, ultra-stable platform consisting nested
and folded Michelson interferometers that can be actively phase
stabilized. This platform generates a square of identical laser
pulses that can be adjusted to have arbitrary time delay between
them, while maintaining phase stability. The JILA-MONSTR provides
output pulses for nonlinear excitation of materials and
phase-stabilized reference pulses for heterodyne detection of the
induced signal. This arrangement is ideal for performing coherent
optical experiments, such as multidimensional Fourier-transform
spectroscopy, which records the phase of the nonlinear signal as a
function of the time delay between several of the excitation
pulses. The resulting multidimensional spectrum is obtained from a
Fourier transform. This spectrum can resolve, separate and isolate
coherent contributions to the light-matter interactions associated
with electronic excitation at optical frequencies. To show the
versatility of the JILA-MONSTR, several demonstrations of
two-dimensional Fourier-transform spectroscopy are presented,
including an example of a phase-cycling scheme that reduces noise.
Also shown is a spectrum that accesses two-quantum coherences,
where all excitation pulses require phase locking for detection of
the signal.
\end{abstract}

\date{\today}
\maketitle

\section{Introduction}

Multidimensional Fourier-transform spectroscopy is an important
tool for elucidating structure and dynamics of
matter.\cite{Ernst1} It was developed at radio frequencies using a
sequence of electromagnetic pulses, although it is applicable to
any region of the electromagnetic spectrum where there exists the
ability to generate, manipulate and detect pulsed radiation. Over
the last decade, multidimensional concepts have been applied in
optical experiments,\cite{Cho} where short intense pulses are
generated by lasers. In the infrared, these techniques can explore
vibronic resonances to determine the structure and coherent
dynamics of molecules.\cite{Hamm1,Asplund,Tokmakoff1,Sul} At
higher optical frequencies the light excites and probes electronic
transitions in molecules\cite{Jonas1,Warren1} and
semiconductors.\cite{Borca,Langbein}

Multidimensional Fourier-transform spectroscopy explicitly tracks
the phase of an induced nonlinear signal generated by the sequence
of excitation pulses.\cite{Mukamel} Tracking the phase with
respect to the time delays between excitation pulses results in a
multidimensional time-domain data set, which is converted to a
multidimensional spectrum with a Fourier-transform. This technique
has several advantages over other electromagnetically driven
spectroscopic techniques: the phase is tracked explicitly,
capturing the induced complex polarization in the sample; coherent
coupling between states is consequently resolved; the nonlinear
response is unfolded onto a multidimensional plane, thus
separating and isolating otherwise overlapping spectral
contributions in the response; and coherent non-radiative
contributions can also be isolated with the appropriate
time-ordering of the excitation pulses.

Numerous methods exist for implementing optical multidimensional
Fourier-transform spectroscopy, all of which require
sub-wavelength stability. These methods can include passive
stabilization, such as sturdy construction and common paths
through optical elements,\cite{Miller1,Fleming1,Selig1} or active
stabilization that employs feedback loops to suppress the effects
mechanical drift.\cite{Joffre1,Hamm2,Zhang1} Previously published
schemes include combinations of diffractive optics elements to
create spatially separated pulses,\cite{Miller1,Fleming1} and
pulse shapers to vary the time and phase of individual
pulses.\cite{Warren1,Nelson1,Zanni1} There are non-collinear,
partially collinear and fully collinear implementations of 2DFT
spectroscopy, the latter of which require phase cycling to extract
the signal.\cite{Warren1,Warren2,Marcus} Most commonly employed is
the background-free transient four-wave mixing (TFWM)
geometry.\cite{Miller1,Fleming1,Hamm2,Zhang1,Selig1,Nelson1}

In this paper, we present an ultra-stable platform that provides
four identical laser pulses in a non-collinear box geometry. This
platform is designed for optical multidimensional
Fourier-transform spectroscopy, and has been dubbed the ``JILA
Multidimensional Optical Nonlinear SpecTRometer" (JILA-MONSTR).
The JILA-MONSTR is based on a three-pulse TFWM excitation scheme,
using the fourth pulse as a phase-stable reference pulse. The
versatility of the platform is demonstrated with several 2DFT
projections, including a technique that extracts two-quantum
coherent and requires all excitation pulses to be phase locked.
Also presented is an optional, straight forward implementation of
phase cycling to reduce noise from scattered pump light in the
non-collinear geometry.\cite{Zanni2} Note that all-optical phase
retrieval is not discussed because extensive details are found in
Ref. \onlinecite{Bristow1}. Semiconductor quantum
wells\cite{Axt,Cundiff1} and potassium vapor\cite{Lorenz} are used
to demonstrate the regular and phase-cycling operation of the
JILA-MONSTR, respectively.

\section{Experimental}
\subsection{Excitation Scheme}

The JILA-MONSTR is based on a three-pulse excitation scheme, which
enables the acquisition of one-, two- and three-dimensional
spectra. The excitation pulses are assigned the notation A*, B and
C, where pulse A* is phase conjugated with respect to pulses B and
C. In the non-collinear box geometry, the excitation pulses lie on
three corners of a square. The TFWM signal therefore has a
phase-matching condition ($k_{sig}$) that is some permutation of
the excitation wavevectors $-k_{A}$, $k_{B}$ and $k_{C}$. Any
permutation of the phase-matching condition can be measured by
changing the time ordering of the pulses, without altering the
signal collection path. Note that two-pulse excitation schemes can
also be recorded in a different direction that require a change in
the signal collection path.

Figure 1 shows the possible pulse sequences available with
three-pulse excitation. The pulse ordering is germane to the
dephasing information induced in the excited material. In
Fig.~1(a) the first and third pulses are phase conjugated with
respect to one another. This excitation sequence undoes the effect
of sample inhomogeneity, leading to a photon echo. Hence, this
pulse sequence is referred to as either $S_{I}$ or ``rephasing."
In this case, because the first and third time axes are
interrogated, the Fourier transform gives a spectrum with respect
to $\omega_{\tau}$ and $\omega_{t}$ (see inset). Figure~1(b) is a
variant of the rephasing projection where the Fourier transform is
with respect to time axes $T$ and $t$, allowing the non-radiative
contributions of the second time period to be
explored.\cite{Yang1} In Fig.~1(c) the conjugate pulse is second
in the sequence and, unlike Fig.~1(a), the inhomogeneity is not
cancelled leading to a free decay. This scheme is therefore
referred to as either $S_{II}$ or ``non-rephasing." Similarly, in
Fig.~1(d) rephasing does not occur because the conjugate pulse is
third in the sequence. Hence, this sequence is known as $S_{III}$.
Since the first two pulses are not phase conjugates of one another
they can coherently add allowing access to two-quantum
information. Hence, the second and third time periods are Fourier
transformed to yield the 2DFT spectrum.

This scheme requires three non-collinear excitation pulses and a
heterodyne reference (Ref) and/or tracer (Tr) pulse. All pulses
must be controlled with sub-cycle precision over delay ranges that
are large compared to the coherence length of the TFWM emission in
most materials. This requirement can be achieved using mechanical
translation stages and sturdy apparatus construction, enhanced by
active feedback electronics. Phase control between any two pulses
can be achieved using the error signal of a co-propagating
continuous-wave (CW) laser beam in a Michelson interferometer. In
which case, the error signal is the interfering term between the
superposition of the CW light in each arm of the interferometer.
Phase control of four pulses can be achieved with a set of nested
interferometers; see the schematic diagram shown in Fig.~2. In
this setup, the pulses exit through the dichroic mirrors (DCMs)
and the CW laser is reflected, producing error signals at the
diagnostic ports. Error signals are recorded and fed into
electronic loop filters that adjust piezo-electric transducers
(PZTs) in each interferometer (not shown in this diagram). Several
delay stages are placed in convenient locations within the setup.
To create the box of output pulses this system must be folded.

\subsection{The JILA-MONSTR}

The scheme presented above is implemented in the JILA-MONSTR,
which comprises two decks with the optics for an interferometer on
each deck and an additional interferometer between the
decks.\cite{Pshenichnikov} Computer-aided design drawings of the
lower and upper decks are shown in Fig.~3(a) and (b),
respectively. The decks are made from cast aluminum and have a
mass of 39~kg each. This material avoids both longer term
relaxation of rolled metal slabs and reduces shorter term thermal
drift. Each deck is milled, using computer numerical control
software, to house the optics for each interferometer. The lower
deck supports the interferometer for pulses C and Ref; the long
delay stage (U), which moves C and Ref together with respect to
the top deck pulses; the PZT for the bottom-deck interferometer;
and the short delay stage (Z), which changes the time delay
between pulses C and Ref. The top deck supports the interferometer
for pulses A* and B; the PZTs to control the phase of the top- and
inter-deck interferometers; and two short delay stages (X and Y),
which allow arbitrary variation of time position of pulse A* and
B. This platform has a total of four direct-drive linear
translation stages with sub-nanometer resolution: X,Y,Z stages
have 5~cm of travel and the U stage has 20~cm of travel.

The decks and all optical mounts are custom made, except for two
sturdy three-axis, kinematic mirror mounts. All mirrors are coated
with protected Ag for operation close to 800~nm. Three broadband
beamsplitters (BS) divide the input laser pulse into four
replicas. The BS coatings are centered at 800~nm, on thin and low
group-velocity dispersion substrates. Compensation plates (CP),
made from identical substrates, balance the dispersion in the
appropriate arm of each interferometer. All transparent and
semitransparent optics are antireflection coated, and mounted in
strain-minimizing mounts with o-rings to prevent induced
birefringence.

The decks form the exoskeleton of the JILA-MONSTR, enveloping the
optics in a 5-cm gap; see Fig.~3(c). Five mounting points ensure a
sturdy enclosure. The small gap reduces exposure of the optics to
air currents, and they are further protected with side panels.
After assembly the 5-cm DCM is attached to the front of the
JILA-MONSTR, reflecting the CW laser light back into the device
and completing the interferometers. The interferometric error
signals are shown as dashed lines emerging from the right-hand
side of the JILA-MONSTR (denoted R in the figure). The laser
pulses that perform the 2DFT experiment emerge through the DCM in
a square that is 5~cm along each side. Figure~4 shows a photograph
of the closed JILA-MONSTR with the input and output beams
highlighted.

Alignment of the JILA-MONSTR starts with the separated decks,
individually prior to assembly. Each deck must provide two
parallel output beams, that do not deviate when the delay stages
are scanned, while producing clean interference patterns at the
diagnostic port. (Only one interferometer works per deck before
closing the assembly). When the assembly is closed, the folded
interferometers are completed by the single 5-cm DCM, which
requires accurate alignment to optimize all three interferometeric
error signals at the diagnostic port. Consequently, the
JILA-MONSTR has been designed such that excess degrees of freedom
are remove from the optical paths. Two quadrant-diode
photodetectors (QD) assist with daily optimization of the
alignment.

Figure~5(a) shows the error signals of the three interferometers
of the JILA-MONSTR, recorded at the diagnostic port with the
active feedback loops disengaged. The path length of the
interferometer arms drift over several minutes, leading to changes
in phase of $>2\pi$. In contrast, the error signals are held at
zero volts when active stabilization is engaged; see Fig.~5(b).
Note that the bottom-deck and inter-deck error signals are offset
by 1~V. The variations of the locked error signals are normal
distributions, with a standard deviation that corresponds to
$\sim2$~nm of motion. This motion corresponds to phase
stabilization of $\lambda / 130$ up to $\lambda / 400$, assuming
that the laser pulses for the 2DFT experiment have a wavelength of
800~nm.

To perform a 2FDT scan, an individual time delay must be moved
many 100's of times with perfectly spaced increments. Each of
these increments is computer controlled. This control process is
discussed in more detail in the next subsection, after the
introduction of the setup downstream from the JILA-MONSTR.

\subsection{Setup}

Figure~6 shows the entire experimental setup for multidimensional
spectroscopy. The pulsed laser source is a mode-locked Ti:sapphire
oscillator. It operates at 76~MHz, producing $\sim$100~fs pulses
that are tuned near 800~nm or to 768~nm for the semiconductor or
potassium vapor experiments, respectively. The CW stabilization
source is a HeNe laser operating at 632.8~nm. These are combined
on a DCM situated before the entrance to the JILA-MONSTR.

The pulses from the JILA-MONSTR are focused by a single lens to
overlap at the sample under investigation, as shown in the
photograph (Fig.4) and schematic diagram (Fig.~6). Three of the
pulses are excitation pulses, generating the TFWM signal. The
fourth pulse is attenuated and is used as the heterodyne reference
pulse. Note that for many 2DFT experiments the reference (local
oscillator) transmits through the sample. In some cases however,
the nonlinear signal is too sensitive to the excitation
conditions. Hence, a phase-stabilized reference is routed around
the sample and recombined with the signal further downstream
(details not shown in Fig.~6).\cite{Zhang1,Bristow1} In either
case, the signal and reference pulses propagate collinearly to the
spectrometer, temporally delayed by several picoseconds. A
spectral interferogram is captured by the spectrometer (see inset
of Fig.~6), from which the complex spectrum is
obtained.\cite{Joffre2}

A 2DFT spectrum is constructed by incrementally adjusting one time
delay while recording spectral interferograms. The complete set of
data is converted to a two-dimensional spectrum by a
one-dimensional numerical Fourier transform along the direction of
the scanned time axis. The incremental adjustment and acquisition
of the 2DFT spectrum is automated, as shown in the flow chart in
Fig.~7(a). The sequence of events is: the feedback loops of the
appropriate interferometers are disengaged; the desired
translation stage(s) are moved a specific distance; after a short
time, allowing the delay stage to reach its new position, the
feedback loops are reengaged; and a spectral interferogram is
recorded. This procedure is the main loop of the 2DFT acquisition
algorithm, and is repeated until the length of the scan is
sufficient to collect the full dephasing of the TFWM signal.
Within the main loop, the locked and unlocked error signals are
recorded every time the feedback loops are disengaged. The
recorded values are used to recalculate the distance to move the
delay stage for that step. Repeating this calculation every step
ensure that the feedback loops are never forced to the extremes of
their compensation range, and speeds up the scan rate. Even though
a timeout can occur before the feedback loops are reengaged, the
scan is sufficiently stable that only a catastrophic electronics
glitch cannot be corrected. While scan rates are high, continuous
scanning methods can potentially improve acquisition rates to
shorten the overall acquisition time.

In addition to recording the error signals and PZT voltages, the
experimenter can monitor the error signals in real time on a a
100~MHz, 4-channel oscilloscope. Figure~7(b) shows a typical
screen capture of one error signal during a single step of
distance commensurate with 4 HeNe fringe. (Nyquist undersampling
is used during the data analysis to correctly determine the
frequency of the 2DFT signal after the Fourier
transform.\cite{Zhang1}) The fast oscillations are due to the
motion of the stage, when the feedback loop is disengaged. The
shaded regions show when the feedback loops are engaged while the
spectral interferogram is recorded.

Two samples are used to demonstrate the versatility of the
JILA-MONSTR. Firstly, data is shown from 4-period multiple quantum
well sample grown by molecular beam epitaxy. The semiconductor
sample consisting of 10~nm GaAs wells and Al$_{0.3}$Ga$_{0.7}$As
barriers, with an etch stop layer that allows chemical removal of
the thick GaAs substrate. Measurements are performed in a
cryostat, cooled to $\sim$6~K. Secondly, results are presented
from a thin potassium vapor cell, where the potassium is heated to
523~K (250~$^\circ$C). The thickness of the cell is approximately
6~$\mu$m. In both samples the effective optical thickness is
chosen to minimize reabsorption of the induced radiation.

\section{Results}

The JILA-MONSTR is capable of performing one-dimensional linear
and non-linear spectroscopy as well as two- and three-dimensional
non-linear spectroscopy. Versatility arises from the pulse layout.
Single-pulse linear measurements can be performed to obtain the
absorption or photoluminescence. Two-pulse nonlinear pump-probe
measurements can also be performed using the attenuated tracer
pulse as a probe and any one of the other pulses as a strong pump.
Three-pulse excitation results in TFWM signals and
higher-dimensional Fourier-transform spectroscopy. In this
section, 2DFT results are presented with references to related
work.

Linear absorbance for the quantum well sample is shown in the top
panels of Fig.~8. Two resonances are observed, which are
associated with bound electron-hole pairs, or exciton states. The
lower and higher energy absorption peaks are the heavy-hole
($X_{hh}$) and light-hole ($X_{lh}$) excitons, respectively.
Increased absorption on the higher energy side of $X_{lh}$ is due
to absorption from the heavy-hole valence band to the continuum
conduction states at energies above the discrete states of the
quantum wells. The binding energy of the $X_{hh}$ is approximately
8~meV, which is similar to the $X_{hh}$ and $X_{lh}$ separation in
this sample.

\subsection{2DFT Spectra}

The complexity of the coherent response of semiconductor quantum
wells lends itself to study by 2DFT spectroscopy, because this
technique separates many of the competing processes. Figure~8 (b,c
and e) shows the real part of the rephasing, non-rephasing and
two-quantum 2DFT spectra. The emission axis defines the numerical
sign of the photon energies. Due to the location of pulse A* in
the excitation sequence, the absorption energy is in the positive
(negative) quadrant for non-rephasing (rephasing). These spectra
are normalized to their respective strongest features.

Spectral features that appear on the diagonal (in the
single-quantum spectra) and double-diagonal (in the two-quantum
spectrum) are coherent intra-action of the $X_{hh}$ or $X_{lh}$.
Off-diagonal features correspond to coherent interaction between
these states. The coherent response of the quantum well is very
sensitive to the excitation conditions, and the spectral features
reveal the presence of many-body interactions.\cite{Li1,Zhang2}
Additionally, elongation of features in the rephasing 2DFT spectra
arises from inhomogeneous broadening due to well width
fluctuations.\cite{Kuznetsova1} Fig.~8(b) and (c) are acquired
with co-circularly polarized pump pulses, which excludes the
formation of biexcitons. Biexcitons are observed for other
polarization configurations and their 2DFT spectroscopic
fingerprint have been discussed recently.\cite{Zhang2,Bristow2}
Biexciton contributions are also observed in the two-quantum 2DFT
spectra.\cite{Yang2,CundiffACR}

The coherent two-quantum transition in semiconductor quantum wells
have been previously studied with TFWM.\cite{2Q} However, these
experiments have left many unresolved questions. Excitation of
two-quantum coherences requires that all excitation pulses be
phase locked,\cite{Nelson2,Karaiskaj1} and the first two pulses
(in a three-pulse scheme) must not be phase conjugated with
respect to one another. For $S_{III}(\tau,\omega_{T},\omega_{t})$
the pulse sequence is B then C followed by A*; see Fig.~1(d). The
final pulse then examines coherent transitions from the
two-quantum to one-quantum states.

Figure~8(d) shows the linear absorption for the quantum well
sample. The excitation laser is tuned over the $X_{hh}$.
Figure~8(e) shows an example of the real part of 2DFT spectrum of
GaAs quantum wells using the two-quantum technique, where $\tau =
0$~fs and the polarization configuration is cross-circular. The
two-quantum spectrum is plotted as a function of two-quantum
absorption energy ($2\hbar\omega_{T}$) versus the emission photon
energy ($\hbar\omega_{t}$). The dashed line marks the
double-diagonal. In the $S_{III}$ spectrum the two-quantum
contributions are many-body scattering states.\cite{CundiffACR}

\subsection{Phase Cycling and Noise Reduction}

Phase cycling has been used in multidimensional NMR\cite{Ernst1}
and collinear optical 2DFT
spectroscopy\cite{Warren1,Warren2,Marcus} to extract signals and
remove noise, interference effects or multiple-quantum
contributions. It is therefore useful for non-collinear
experiments\cite{Zanni2} that suffer from a low signal-to-noise
ratio. Some situations may be dominated by noise sources such as
pump scatter, which emanates from the same location as the TFWM
signal. In 2DFT spectra, pump scatter is observed along the
diagonal of the single-quantum spectra, because the pump is only
self-coherent, i.e. only correlated when
$\omega_{\tau}=\omega_{\rm t}$. This is not an issue for
two-quantum spectra, because the Fourier transform yields a
different spectral range, without pump scatter. In the
single-quantum 2DFT spectra, cycling the phase of the excitation
pulses during a 2DFT scan suppresses the noise along the diagonal.

In NMR spectroscopy phase alternative pulse sequences (PAPS) are
used, where the relative phase of the two RF pulses switches
between in phase and out of phase every time step in a scan.
Incorrectly phased artifacts in the detected signal then cancel
when the Fourier transform is performed, leaving only the signal
in the two-dimensional spectrum. The optical analogy of PAPS
employed here records and averages two spectral interferograms
that have nearly identical time delay and a controlled phase shift
between two pairs of beams, thus allowing the incorrectly phased
noise to cancel when averaged. The example used here is for an
adjustment of the second time period $T$, such that the phase that
is cycles is proportional to $\phi_{T}=n\pi+\omega_{T}T$, where
the first term is toggled between 0$\pi$ and 1$\pi$ for every time
step along axis $\tau$. Figure~9(a) shows how the phase-cycling
scheme is implemented in the four pulses excitation sequence of
the JILA-MONSTR. The phase between the first two pulses (A* and B)
and the second two pulses (C and Ref) are shifted. In the
JILA-MONSTR, this is achieved by moving the long delay stage (U)
back and forth for every incremental position of the first
excitation pulse; see the inset of Fig.~9(b). The motion uses the
same disengage-move-reengage algorithm shown in the main loop of
Fig.~6(a). In practice the wavelengths of the excitation and
phase-control lasers are incommensurate: the example presented
here is performed on the D$_{\rm 1}$ and D$_{\rm 2}$ lines of
potassium vapor; thus, the Ti:sapphire excitation wavelength is
768~nm. The most convenient step size of the U stage is therefore
6 HeNe fringes, which corresponds to approximately 5$\pi$ at the
Ti:sapphire wavelength.

Figure~9(b) shows an example of the TFWM signal plotted along the
reconstructed emission time $t$, with and without phase cycling.
The reconstructed axis is obtained by the numerical Fourier
transform of the heterodyne-detected interferogram. In these
transient the time delays $\tau$ and $T$ are set to 1~ps. The
dashed transient is acquired for normal operation of the
JILA-MONSTR, and thus constructed from a single interferogram at
0$\pi$. In contrast, the averaging of the phase-cycled spectral
interferograms is shown as a solid line, where contributions from
A* and B are suppressed.

Figure~10 shows 2DFT spectra for both normal and phase-cycling
operation of the JILA-MONSTR. The top (bottom) row shows
non-rephasing (rephasing) data, and the left column is acquired
without phase cycling, where diagonal streaks are seen in both
spectra. These diagonal streaks are suppressed in the right-hand
side of the figure, showing the advantage of phase cycling and the
rapid temporal flexibility of the JILA-MONSTR. Note that more
complex phase-cycling schemes can be employed to remove the effect
of all pump pulses.

\section{Conclusion}

In summary, we have shown a versatile platform for performing
optical multidimensinoal Fourier-transform spectroscopy. The
JILA-MONSTR uses an active stabilization scheme to provide four
identical phase-locked pulses arranged in a box geometry. The
active, computer-controllable feedback loops combined with
electromechanical delay stages allow the phase-locked pulses to be
ordered in any sequence with arbitrary time delays. The
JILA-MONSTR has a stability of greater than $\lambda$/100 and
employs a stepping scheme that allows for acquisition of full 2DFT
spectra, even for samples with long dephasing times.

The JILA-MONSTR can perform one-, two- and three-dimensional
spectroscopy. It can be used for numerous excitation techniques
within the scope of the various dimensions. Here we have
demonstrated selected examples of $S_{I}$, $S_{II}$ and $S_{III}$
techniques. In addition to the versatility, stability and speed of
acquisition, the JILA-MONSTR offers straightforward implementation
of phase-cycling to improving data quality. Here we have
demonstrated a useful phase-cycling scheme to suppress pump
scatter.

The JILA-MONSTR has proven to be a robust platform for producing
multidimensional spectra. It is a viable way to separate and
isolate the competing contributions to coherent light-matter
interactions. This sophisticated device has great potential in the
future of optical spectroscopy.

\bigskip

Acknowledgements: The authors thank Richard Mirin for the
epitaxially grown quantum well sample. This work was supported by
the National Science Foundation and the Chemical Sciences,
Geosciences, and Biosciences Division Office of Basic Energy
Sciences, U.S. Department of Energy.

\newpage

\begin{figure}[t]
\centering{\includegraphics[width=8.0cm]{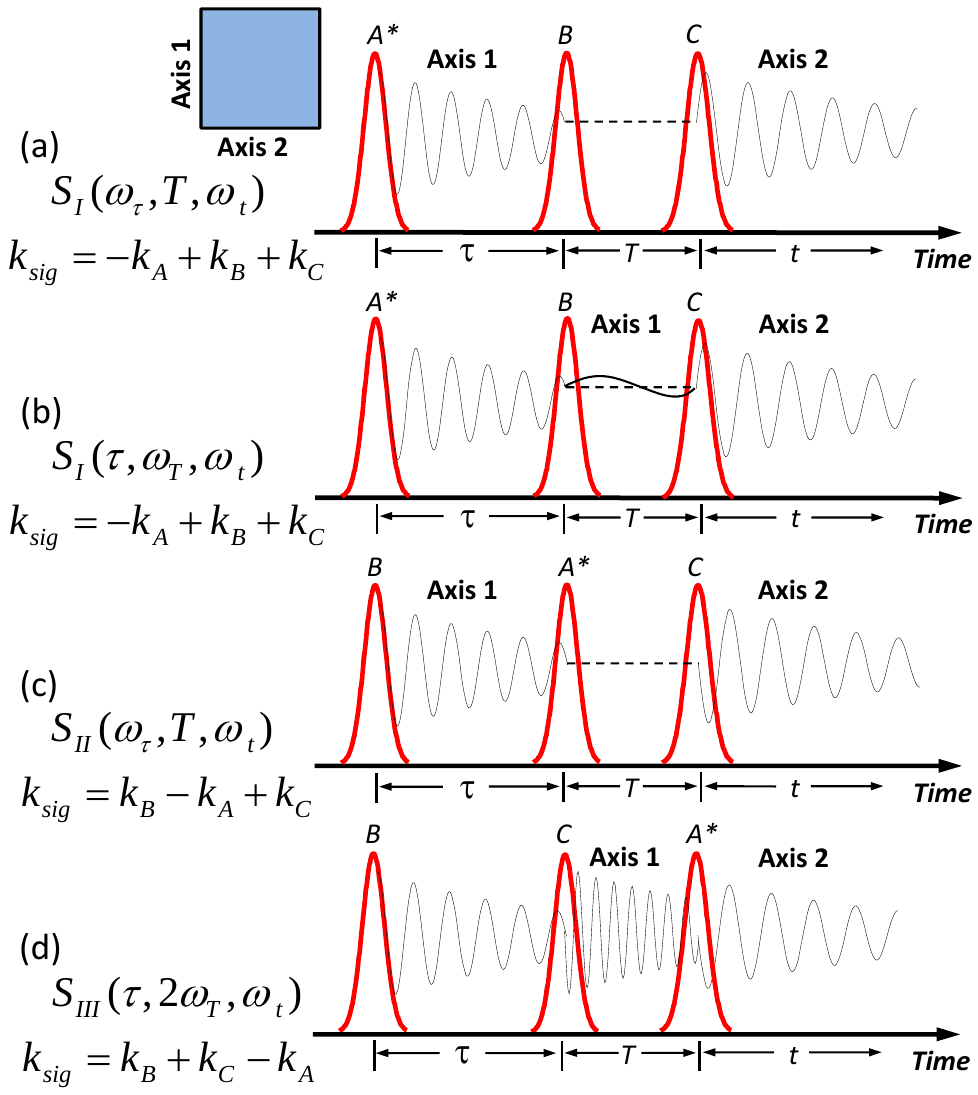}} \caption{(Color
online) Schematic diagrams of the pulse ordering of optical 2DFT
techniques available with a three-pulse excitation scheme: (a)
represents the ``rephasing" $S_{I}$ technique; (b) is a variant of
(a) where the middle time axis is scanned to examine non-radiative
contributions; (c) is ``non-rephasing $S_{II}$ technique; and (d)
is the two-quantum $S_{III}$ technique. In each case Axis 1 and
Axis 2 are labelled for comparison to the inset of (a) showing
which axes are plotted in the resulting 2DFT spectrum.}
\label{fig1}
\end{figure}

\newpage

\begin{figure}[t]
\centering{\includegraphics[width=8.0cm]{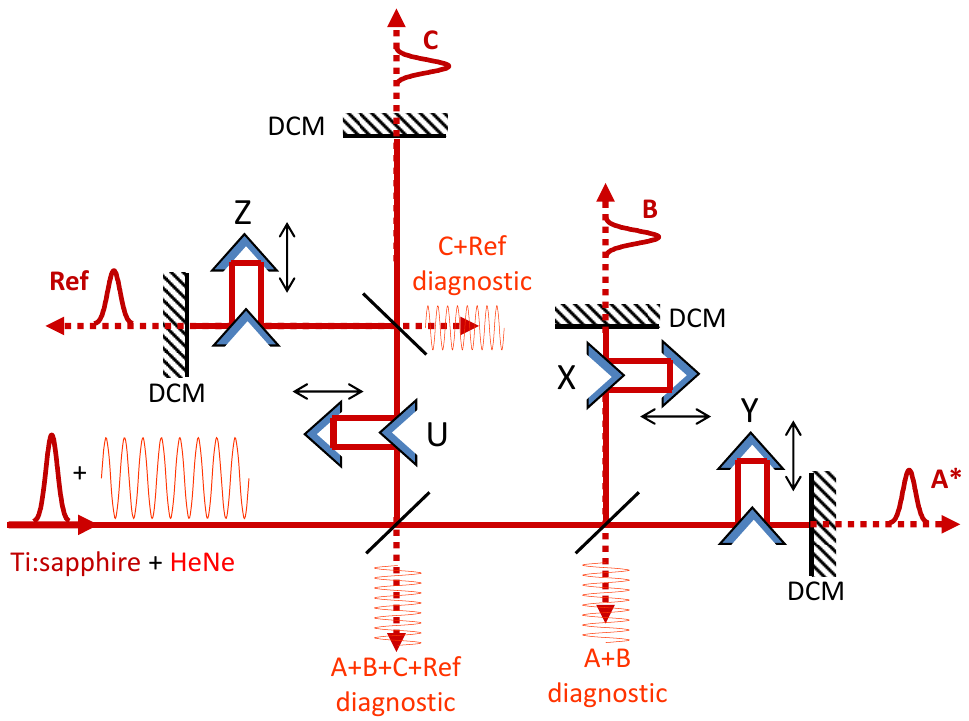}} \caption{(Color
online) Schematic representation of the nested Michelson
interferometer layout. A*, B, C and Ref are the Ti:sapphire laser
pulses used to perform the 2DFT experiment. The HeNe laser beam is
reflected from the DCM to provide diagnostic information for the
interferometers. The proposed layout is folded such that the DCM
is a common optical element.} \label{fig2}
\end{figure}

\newpage

\begin{figure*}[t]
\centering{\includegraphics[width=9.5cm]{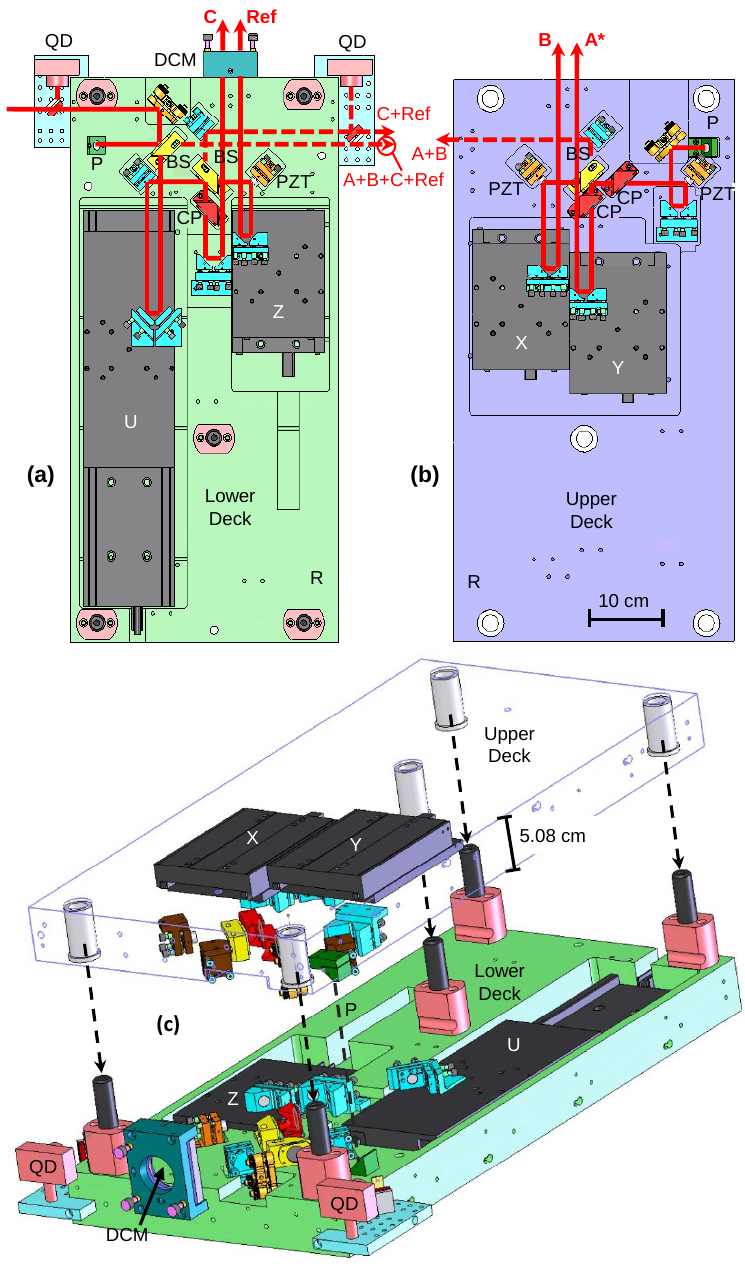}} \caption{(Color
online) Computer Aided Design drawings of the JILA-MONSTR's (a)
lower and (b) upper interferometer decks. P refers to the
periscope between decks and R indicates the right-hand side on
each deck. Laser input is on the left side of the lower deck and
the Ti:sapphire pulses emerge from the front through the dichroic
mirror (DCM), two from each deck. Error signals for each
interferometer emerge from the right-hand side as dashed lines,
two from the bottom and one from the top decks. [BS =
beamsplitter, CP = compensation plate, PZT = piezo-electric
transducer, QD = quad-diode photodetector.] (c) Three-dimensional
CAD drawings of the entire assembly showing how the two decks are
assembled.} \label{fig3}
\end{figure*}

\newpage

\begin{figure}[t]
\centering{\includegraphics[width=8.0cm]{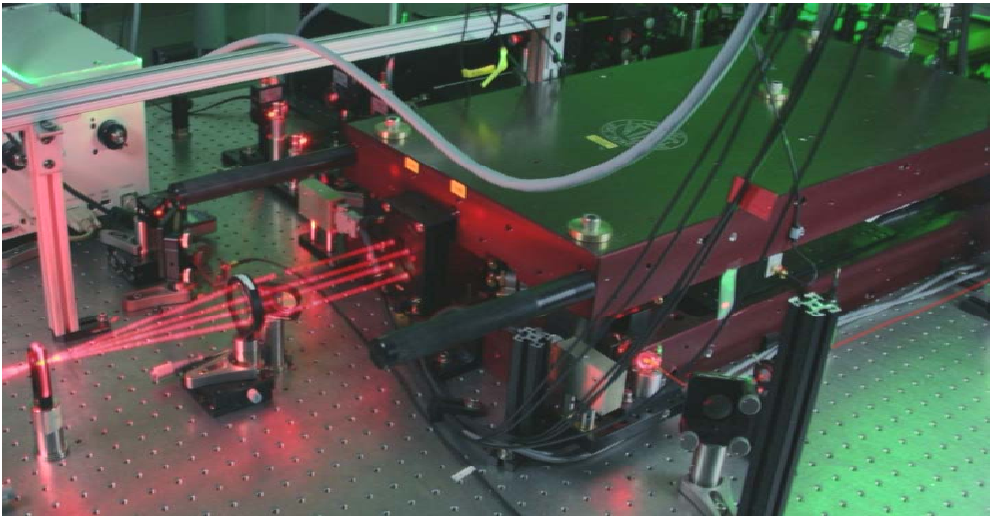}} \caption{(Color
online) Photograph of the JILA-MONSTR highlighting the input beam
on the right-hand side of the picture and four output beams
focused to a single location.} \label{fig4}
\end{figure}

\newpage

\begin{figure}[t]
\centering{\includegraphics[width=7.5cm]{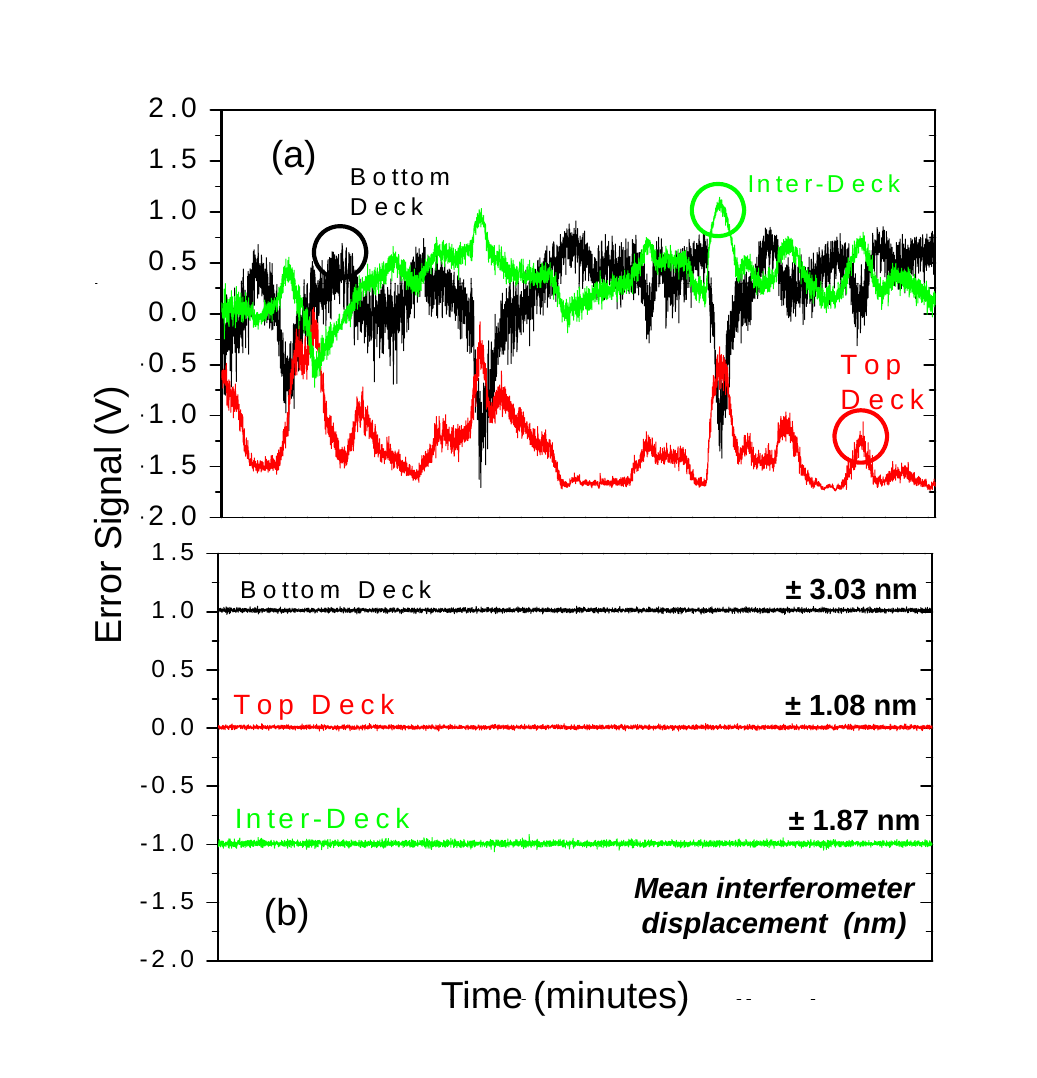}} \caption{(Color
online) The error signals for the top-deck, bottom-deck and
inter-deck interferometers recorded for approximately 10 minutes
(a) without and (b) with active stabilization engaged. In (a)
circles link each trace to the appropriate label. In (b) the
bottom-deck and inter-deck signals are offset from zero for
presentation purposes.} \label{fig5}
\end{figure}

\newpage

\begin{figure}[t]
\centering{\includegraphics[width=8.0cm]{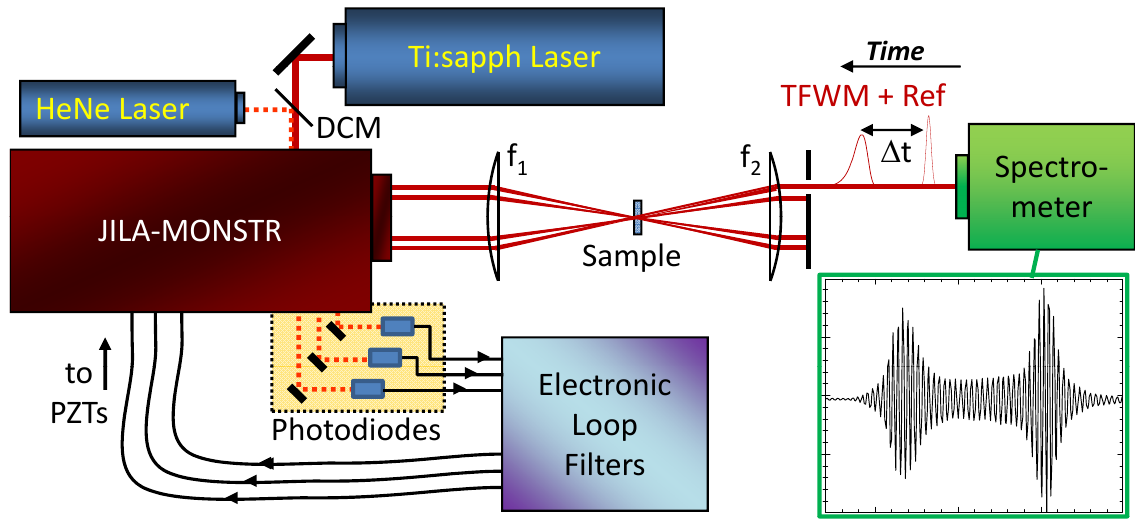}} \caption{(Color
online) Experimental setup for 2DFT spectroscopy showing the pump
and phase-stabilization laser, the JILA-MONSTR with diagnostic
port, and feedback loops. Also shown is the arrangement of the
sample and the collection of light in the spectrometer. A
heterodyne reference pulse is used to perform spectral
interferometry and reconstruct the time axis t. The inset shows a
typical spectral interference pattern recorded for the
quantum-well sample.} \label{fig6}
\end{figure}

\newpage

\begin{figure}[b]
\centering{\includegraphics[width=8.0cm]{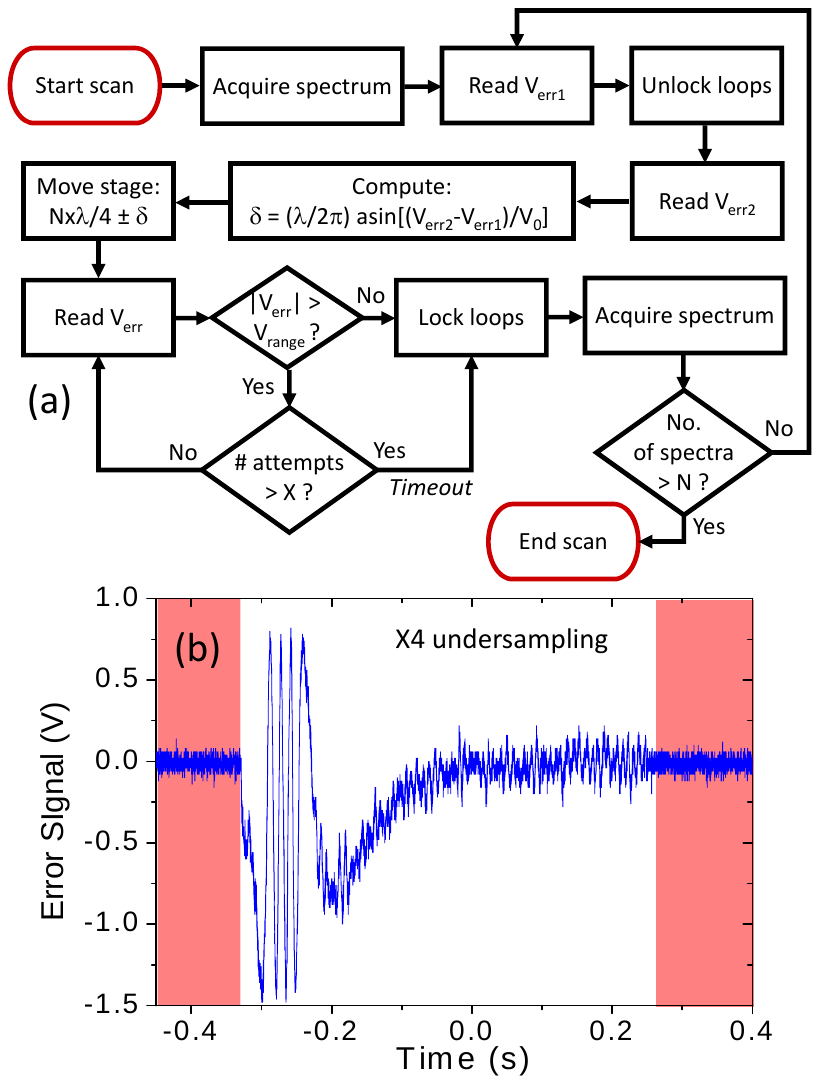}} \caption{(Color
online) (a) Flow chart of the stepping algorithm to ensure
identical steps. There are several measurements of the error
signal: V$_{\rm err1(2)}$ occur before(after) unlocking the loop
filter and V$_{\rm err}$ is measured after the stage has been
moved. V$_{0}$ is half the maximum peak-to-peak error signal, X is
maximum attempts to wait for the error signal to obtain a value
within V$_{\rm range}$ (near zero), and N is the total number of
spectrum acquired in the 2DFT scan. (b) A typical screen capture
of the oscilloscope that monitors the error for the top-deck
interferometer while it is being scanned 4 HeNe fringes. The
shaded regions mark when the the feedback loop is engaged.}
\label{fig7}
\end{figure}

\newpage

\begin{figure}[t]
\centering{\includegraphics[width=8.6cm]{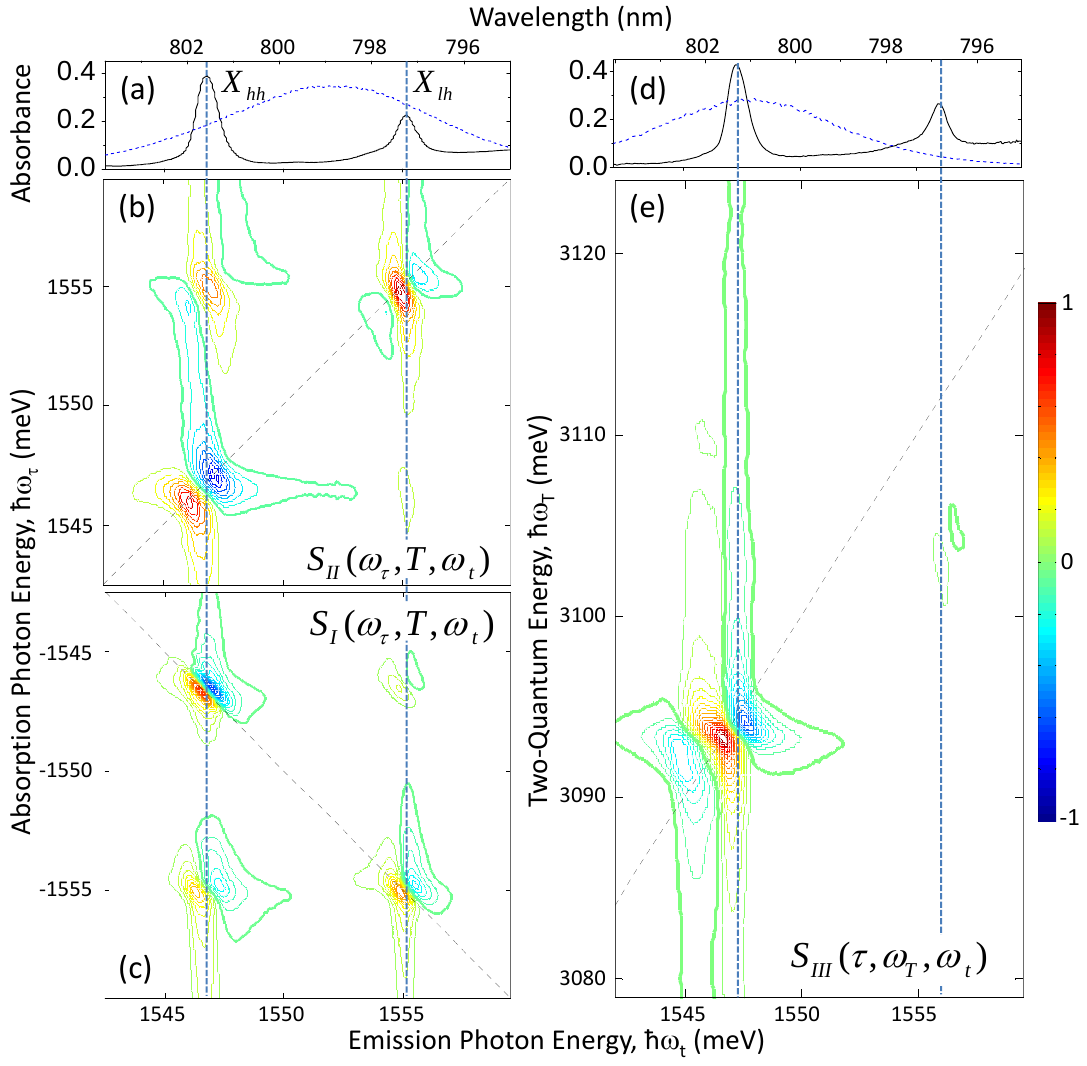}} \caption{(Color
online) (a,d) Linear absorbance spectrum of the GaAs quantum well
sample. Dashed lines show the excitation laser spectra.
Two-dimensional Fourier-transform spectra of GaAs quantum wells,
showing (b) non-rephasing ($S_{II}$), (c) rephasing ($S_{I}$) and
(e) two-quantum $S_{III}$ techniques. The 2DFT spectra are
normalized to the strongest feature, and the thicker contour line
encloses the regions of the negative signal. All 2DFT spectra are
for $T = 0$~fs excitation. The single- (b,c) and two-quantum(e)
spectra are acquired for co-circular and cross-circular
excitation, respectively.} \label{fig8}
\end{figure}

\newpage

\begin{figure}[t]
\centering{\includegraphics[width=7.0cm]{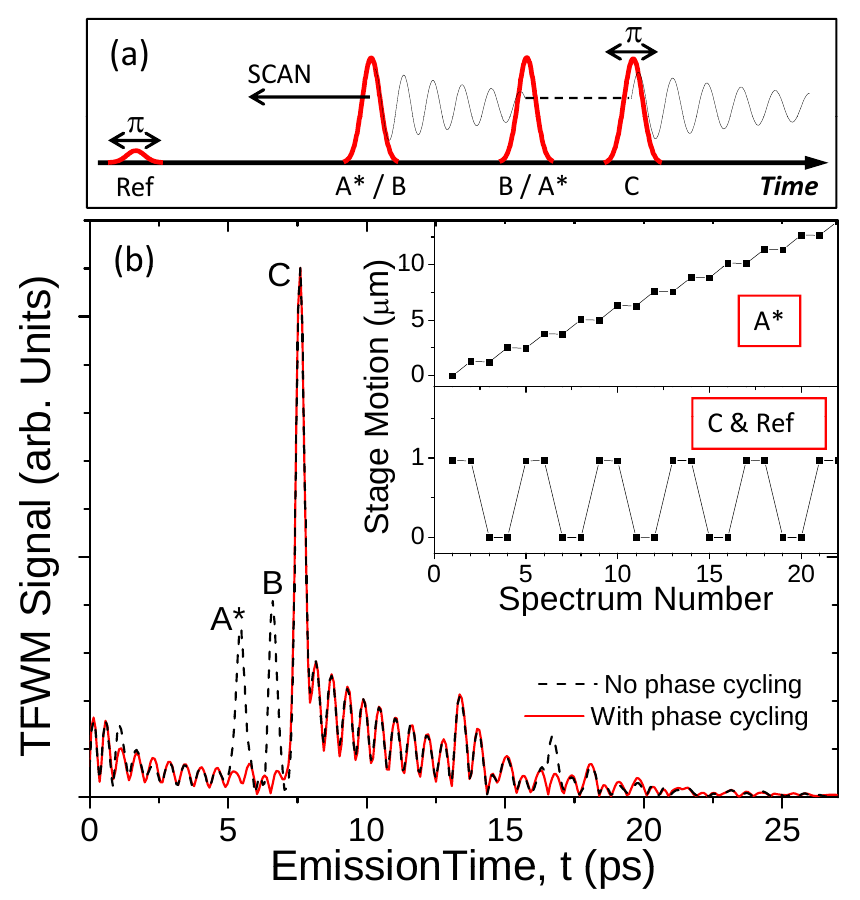}} \caption{(Color
online) (a) Schematic method of the phase cycling operation for
2DFT spectroscopy. While a time delay is scanned, pulses C and Ref
are simultaneously toggled back and forth by $\pi$ and two
spectral interferograms are recorded per data point. (b) Shows the
fourier transform of recorded spectral interferograms taken when
$T$ and $\tau$ are approximately 1~ps. The dashed line is a result
of a single interferogram (i.e. normal operation) and the solid
line is for the difference of the two interferograms. The inset
shows the stepped slope of the scanned pulse, A*, and the toggled
pulses, C and Ref.} \label{fig9}
\end{figure}

\newpage

\begin{figure}[b]
\centering{\includegraphics[width=8.0cm]{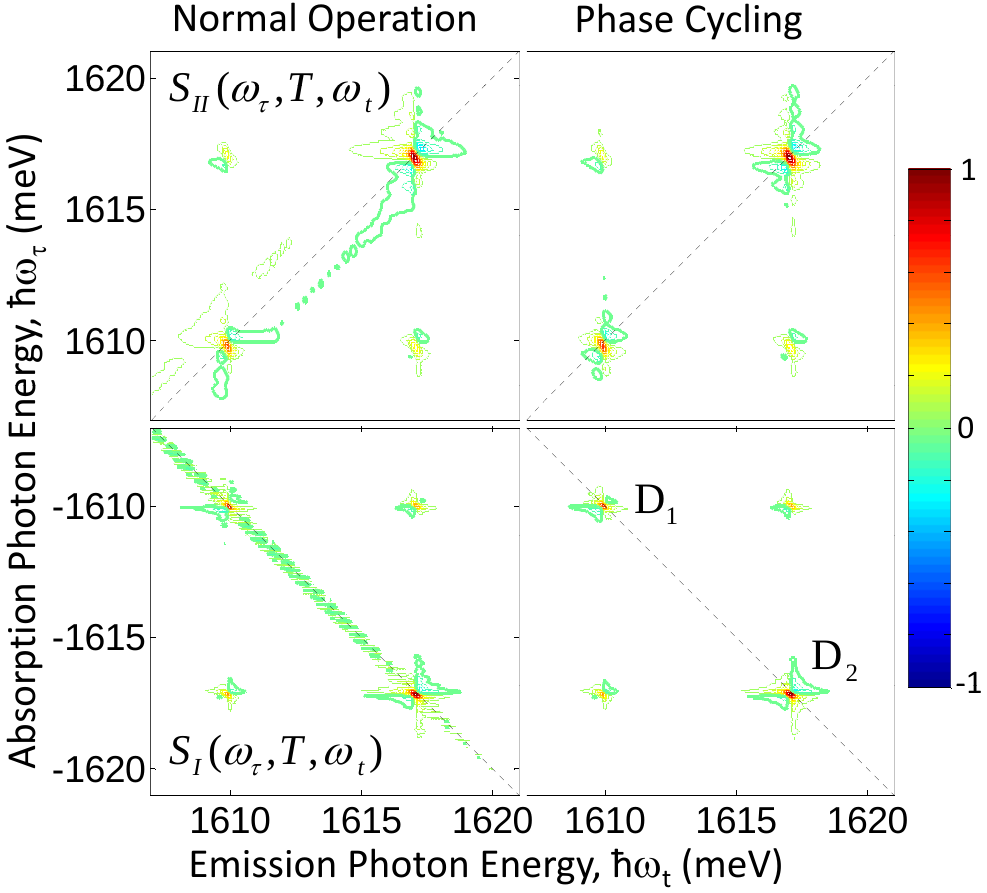}} \caption{(Color
online) An example of noise reduction in the 2DFT spectra of K
vapor using phase cycling. The left-hand panels show rephasing
($S_{I}$) and non-rephasing ($S_{II}$) spectra obtained through
normal operation, and the right-hand panels use phase cycling of
the C and Ref pulses by 6 HeNe fringes.} \label{fig10}
\end{figure}

\end{document}